\documentclass[12pt]{amsart}
\usepackage{graphicx}

\usepackage{amsmath,amsthm}
\makeatletter
\def\amsbb{\use@mathgroup \M@U \symAMSb}
\makeatother

\usepackage{bbold}

\usepackage{amssymb}
\usepackage{caption}
\usepackage[latin1]{inputenc}
\usepackage{enumerate}
\usepackage{mathrsfs}  
\usepackage{dsfont}

\usepackage[bbgreekl]{mathbbol} 

\usepackage{geometry}
\theoremstyle{plain}

\theoremstyle{definition}
\newtheorem*{defi}{Definition}

\newtheorem*{obs}{Remark}

\numberwithin{equation}{section}

\newcommand{\DP}[2]{\frac{\partial\, #1}{\partial\, #2}}

\def\C{\mathcal {C}}
\def\A{{\mathcal{C}}^\infty(M)}
\def\AT{{\mathcal{C}}^\infty(TM)}
\def\AC{{\mathcal{C}}^\infty(T^*M)}
\def\aA{\mathcal{A}}

\def\U{\mathcal {U}}

\def\R{\amsbb{R}}

\def\Ddelta{\text{\Large $\mathbb{\Delta}$}} 

\begin{document}

\title[On the quantization of mechanical systems]
 {On the quantization of mechanical systems}

\author{ J.  Mu\~{n}oz-D{\'\i}az and R. J.  Alonso-Blanco}

\address{Departamento de Matem\'{a}ticas, Universidad de Salamanca, Plaza de la Merced 1-4, E-37008 Salamanca,  Spain.}
\email{clint@usal.es, ricardo@usal.es}

\begin{abstract}
We show what seems to be the key for quantization of classical systems. Given a manifold $M$, each riemannian metric (nondegenerate, of arbitrary signature) canonically determines a quantization rule or ``Correspondence Principle'', which assigns to each classical magnitude  (function in $TM$, subject to certain conditions) a differential operator in $\A$. The issue about the order in which the $p$' and $q$'  are to be taken in quantization  loses all meaning, once the general rule has been fixed. Specified the Correspondence Principle, each ``classical state'' of the system, understood as a vector field on $M$, determines a wave equation for each magnitude. The Schrödinger equation is a particular example of these wave equations.

\end{abstract}
\bigskip

\maketitle

\tableofcontents

\subsection*{Physical magnitudes}\label{magnitudes}

Let $M$ be a smooth  manifold of dimension $n$, $TM$ be its tangent bundle and $T^*M$ be its cotangent bundle; let $\A$ denote the ring of complex valued (infinitely) differentiable functions in $M$.

On $T^*M$ it is defined the Liouvile form $\theta$, by $\theta_\alpha=\alpha$, for each $\alpha\in T^*M$, and where the equality is understood assuming that $\alpha$ is lifted by pull-back from $M$ to $T^*M$. In local coordinates $(x^1,\dots,x^n)$ for $M$ and the corresponding ones $(x^1,\dots,x^n,p_1,\dots,p_n)$ for $T^*M$, the expression of the Liouville form is $\theta=p_j\,dx^j$ (summation with respect repeated indexes is assumed). The 2-form $\omega_2:=d\theta$ is the symplectic form in $T^*M$; its expression in local coordinates is $\omega_2=dp_j\wedge\,dx^j$.

The  interior product with $\omega_2$ establishes an isomorphism of $\C^\infty(T^*M)$-modules between tangent vector fields on $T^*M$ and differential 1-forms on $T^*M$: $D\mapsto i_D\omega_2$. In this isomorphism, the vertical tangent fields (those that, as derivations, annihilate the subring $\A$ of $\AC$) correspond  to the horizontal 1-forms (those 1-forms annihilated, by interior product, by vertical fields).

In local coordinates, the vertical tangent field $\partial/\partial p_j$ is applied on $dx^j$.

The isomorphism established by the symplectic form between the $\AC$-modules of vertical fields  and horizontal 1-forms, is naturally extended to an isomorphism between the tensor algebra of the ``vertical'' contravariant tensor and the ``horizontal'' covariant tensor algebras.

For our present purpose we are not interested in all this tensors, but only those corresponding to symmetrical covariant tensors on $M$: let $\aA$ be the $\A$-algebra of symmetric covariant tensors on $M$, considered (after pull-back) as covariant tensors on $T^*M$; the symplectic form made to $\aA$ a $\A$-algebra $\aA^*$ of symmetric contravariant tensor fields on $T^*M$.
In local coordinates, the polynomial $P(dx^1,\dots,dx^n)$ with coefficients in $\A$, corresponds to the polynomial with the same coefficients, $P(\partial/\partial p_1,\dots,\partial/\partial p_n)\in\aA^*$.

In the tangent bundle the notions of vertical tangent field and horizontal 1-forms are analogous to those we just have considered in $T^*M$. In local coordinates $(x^1,\dots,x^n,\dot x^1,\dots,\dot x^n)$, the horizontal 1-forms are the linear combinations of the $dx^j$ with coefficients in $\AT$ and vertical tangent fields are linear combinations of the $\partial/\partial\dot x^j$. But here, if no additional structure to that of manifold is given, we have no symplectic form. The realization of the tensor algebra $\aA$ that, in $T^*M$, was done an algebra of vertical differential operators, in $TM$ is given as an algebra of functions as follows: for each function $f\in\A$ let us denote by $\dot f$ the function on $TM$ defined by the rule $\dot f(v):=v(f)$, for each $v\in TM$ considered as a derivation $v\colon\A\to\R$; the function $\dot f$ is, essentially, $df$. In general, for each horizontal  1-form $\alpha$ on $TM$  it is defined the function $\dot\alpha$ by $\dot\alpha(v):=\langle\alpha,v\rangle$ (duality). By means of this rule, each tensor $a\in\aA$ define a function $\underline a\in\AT$, that is polynomial in the fibres (a polynomial in the $\dot x$' with coefficients in $\A$). The function $\underline a$ is, essentially, the same as the tensor $a$. In conservative mechanical systems, which are now our subject of study, all the magnitudes appearing in the space of position-velocity states are of this type. For that, we will call \emph{classical magnitudes} to these functions $\underline a$; the ring of the classical magnitudes is, then the ring $\underline\aA\simeq\aA$ obtained by associating to each  tensor $a\in \aA$ the function $\underline a$. In local coordinates, $\underline a(\dot x^1,\dots,\dot x^n)$ is obtained  if $dx^j$ is replaced by $\dot x^j$ in the polynomial $a(dx^1,\dots,dx^n)$.
 As $\A$-algebras we have isomorphisms
 $$\underline\aA\simeq\aA\simeq\aA^*,$$
 in such a way that each classical magnitude $\underline a$ (function on $TM$) is, substantially, the same object that
 (is canonically identified to) the vertical differential operator $a^*$ on $T^*M$.

 It is convenient to establish  this correspondence by a different way, by using the Fourier transform, as follows:

 The Liouville form can be interpreted as a function on the fibred product $TM\times_M T^*M$, by assigning to each vector $v_x\in T_xM$ and each 1-form $\alpha_x\in T^*_xM$ the value
 $$\theta(v_x,\alpha_x):=\langle v_x,\alpha_x\rangle.$$
 When the ``dimensions'' are introduced for the classical magnitudes, $\theta$ has dimension of ``action''. For this reason, in order to give full sense to the transcendent functions of $\theta$ and they have a meaning independent of the measures unities, it must be introduced a constant $\hbar$ with dimension of ``action'' and take $\theta/\hbar$ instead of $\theta$. With this, it makes sense the function $\textrm{exp}\,(i\theta/\hbar)$ on  $TM\times_M T^*M$.

 To specify, let us take as base space for the Fourier transform $\mathcal{S}(TM)$, the set of complex valued functions defined on $TM$, that on each fibre $T_xM$  are $\C^\infty$ and rapidly decreasing they and all their derivatives. Analogous meaning for $\mathcal{S}(T^*M)$.

 In each fibre $T_xM\simeq\R^n$, the usual measure is, up to a constant factor, the unique translation invariant (Haar measure of the group $\R^n$). Once chosen on each fibre this Haar measure, we can define the Fourier transform fibre to fibre:
 $$\mathcal{F}\colon\mathcal{S}(TM)\to\mathcal{S}(T^*M),\quad f\mapsto\mathcal{F}f,$$
 where
 $$(\mathcal{F}f)(\alpha_x):=\int_{T_xM}f(v_x)e^{\frac i\hbar\theta(v_x,\alpha_x)}d\mu(v_x),$$
 being $d\mu$ the Haar measure fixed on $T_xM$.

 By taking local coordinates $(x^1,\dots,x^n)$ on an open set $\U$ of $M$ and the corresponding ones $(x^1,\dots,x^n,\dot x^1,\dots,\dot x^n)$ on $T\U$, $(x^1,\dots,x^n,p_1,\dots,p_n)$ in $T^*M$, we have $\theta(\dot x,p)=p_j\dot x^j$ and then:

$$(\mathcal{F}f)(p):=\lambda_x\int_{T_xM}f(\dot x)e^{{\frac i\hbar p_j\dot x^j}}d\dot x^1\cdots d\dot x^n,$$
where $\lambda_x$ is the constant that fixes the choice of the measure.

By derivation under the integral sign its is obtained the classical formula:
$$\mathcal{F}(\underline a(x,\dot x)f(x,\dot x))=\underline a(x,-i\hbar\partial/\partial{p})(\mathcal{F}f)(x,p)$$
for any function $\underline a$ on $TM$ that is polynomial in the $\dot x$, that is to say, for each symmetric covariant tensor $a$ on $M$.

We see that, by changing the symplectic form $\omega_2$ by $(i/\hbar)\omega_2$, the correspondence between classical magnitudes $\underline a\in\underline\aA$ and vertical differential operators on $T^*M$ is the same that the given by the Fourier transform.
 By introducing already the factor $i/\hbar$, let us denote by $A$ the vertical differential operator on $T^*M$ that corresponds to the magnitude $\underline a$:
 $$\underline a(x,\dot x)\leftrightarrow a(x,dx)\leftrightarrow A(x,\partial/\partial p)=\underline a(x,-i\hbar\partial/\partial p).$$
 $\underline a$ and $A$ are the expresions on $TM$, $T^*M$ of the same object, the tensor $a$.

\subsection*{Introducing a metric on $M$. Quantization}\label{MetricQuantization}
When the structure of smooth manifold is the only one given on $M$, there is no correspondence between points of $TM$ and points of $T^*M$, although the symplectic structure (or the Fourier transform fiberwise) has allowed us to establish the correspondence $\underline a\to A$ between functions on $TM$ which are polynomial on fibers and vertical differential operators on $T^*M$.

Let $T_2$ be a riemannian metric (of arbitrary signature) given on $M$. The metric establishes an isomorphism of fiber bundles $TM\to T^*M$ by assigning to each tangent vector $v_x\in T_xM$ the differential 1-form $\alpha_x:=i_{v_x}T_2\in T^*_xM$ (interior product of $v_x$ with $T_2$). By means of that isomorphism we can translate each structure from one to the other of those bundles; we can talk about the Liouville form, the symplectic form, etc., on $TM$. In other to simplify the notation, if there is no risk of confusion, we will use the same notation for each object on $TM$ and its translation to $T^*M$. In this way, if the local coordinated expression for the metric is
 $T_2=g_{ij}dx^i\,dx^j$, the function $p_k$ on $T^*M$ is, in the coordinates of $TM$, $p_k=g_{jk}\dot x^j$, and
 $$\frac{\partial}{\partial {p_k}}=g^{kj}\frac{\partial}{\partial{\dot x^j}}.$$

 In the previously established correspondence $\underline a\to A$, to the function $\dot x^j$ there corresponds $-i\hbar\partial/\partial p_j$; therefore, to the polynomial $p_\ell=g_{\ell j}\dot x^j$ it corresponds
 $$-i\hbar\, g_{\ell j}\,\DP{}{p_j}=-i\hbar\, g_{\ell j}g^{jk}\,\DP{}{\dot x^k}=-i\hbar\,\DP{}{\dot x^\ell}.$$

To the very metric tensor $T_2$ there corresponds on $TM$ the function $2T=g_{jk}\dot x^i\dot x^j$ and, on $T^*M$, the differential operator
 $\text{\Large ${\Ddelta}$}=-\hbar^2g_{k\ell}\partial/\partial p_k\partial/\partial p_\ell$, which, once introduced the metric on $M$ becomes
$$\text{\Large ${\Ddelta}$}=-\hbar^2g_{k\ell}g^{kr}g^{\ell s}\,\DP{}{\dot x^k}\DP{}{\dot x^\ell}=-\hbar^2g^{rs}\DP{}{\dot x^r}\DP{}{\dot x^s}.$$

For each point $x_0\in M$ the metric $T_2$ establishes a local isomorphism between a certain neighborhood $\U$ of the origin in $T_{x_0}M$ and a neighborhood $U$ of $x_0$ in $M$:
$$\textrm{exp}\colon \U\to U$$
which sends each vector $v_{x_0}\in T_{x_0}M$ to the point $\textrm{exp}(v_{x_0})$ which is the end point of the arc of geodesic path in $M$ parameterized by $[0,1]$ and starting at $x_0$ with tangent vector $v_{x_0}$.

By taking local coordinates $(x^j)$ on an neighborhood of $x_0$ in $M$ and the corresponding ones  $(x^j,\dot x^j)$ in $TM$, the expression of $\textrm{exp}$ is
$$x^j=x_0^j+\dot x^j-\frac 12\Gamma_{rs}^j(x_0)\dot x^r\dot x^s+\cdots$$
(where the $\Gamma$' are the Christoffel symbols of the metric), if we get the Taylor expansion till second order terms. That formula is easily derived from the differential equation defining the geodesics and, from it and the inverse function theorem it results that $\textrm{exp}$ is a local differentiable isomorphim. For further details, see \cite{Eisenhart}.

The exponential map allows us to assign to each function $f\in \A$ a function $\widehat f$, defined on a neighborhood of the $0$ section of $TM$, by means of the rule
$\widehat f(v_x):=f(\textrm{exp}\,v_x)$, for the $v_x\in T_xM$ on which the exponential map is defined. The function $\widehat f$ is the description of $f$ done from each point of the configuration space (from ``each observer''). If we denote by $\mathcal{O}(M)$ the ring of germs of $\C^\infty$ functions on neighborhoods of 0 section of $TM$, the assignation $f\to\widehat f$ determines an injection of $\A$ into $\mathcal{O}(M)$ that  we will  call the \emph{riemannian injection}.

The trivial injection $\A\to\C^\infty(TM)$ given by the pull-back associated with the projection $TM\to M$, translates all vertical differential operators in $TM$ that without 0-order terms, to the identically $0$ operator in $\A$. But, thanks to the riemannian immersion, each differential operator $A$ in $TM$ gives on $M$ a non trivial differential operator:
$$\widehat a\,f:=\left.\left( A\widehat f\right)\right|_{0\,\, \text{section}},$$
by identifying $M$ with the 0-section of $TM$.

And this seems to be the key of the quantification:
\begin{defi}
Let $a$ be a symmetric covariant tensor on $M$, $\underline a\in\C^\infty(TM)$ its associated function (``classical magnitude'') and $A$ the vertical differential operator in $TM$ corresponding to $\underline a$ (once identified $TM$ and $T^*M$ by means of the metric). The differential operator
$$\widehat a\colon\A\to\A$$
derived from $A$ by means of the riemannian injection $\A\to\mathcal{O}(M)$ is the \emph{quantification of the magnitude $\underline a$}.
\end{defi}

The quantification $\underline a\to\widehat a$ is $\A$-linear for the module structure in the set of differential operators (given by left multiplication by functions). It is also injective, since, if the vertical differential operator $A$ annihilates all $\widehat f$ coming from $\A$, it holds $A=0$, because on each fibre of $TM$, $A$ is a polynomial in the $\partial/\partial x$ with constant coefficients. However, the multiplicative structure changes, so losing the commutativity: given two symmetric covariant tensors $a$, $b$ in $M$, for each $f\in\A$, the computation of $\widehat b(\widehat a\,f)$   is made by applying the differential operator $B$ to the function $\widehat{\widehat af}$, that differs from $A\widehat f$; this is why, in general, it does not hold $\widehat b\circ\widehat a=\widehat{ba}$. In general neither is true $\widehat{a^2}=\widehat a^2$, as we will see later in some particular instance.

For the time being, the Taylor expansion of second order of the exponential map, allows us to find the quantum operators corresponding to the functions that are polynomials of degree lower or equal than 2 in the $\dot x$.

For each $f\in\A$ we have
$$\left(\DP{\widehat f}{\dot x^k}\right)(0)=\left(\DP{f}{x^k}\right)(x_0),\quad\text{and hence,}\quad \widehat p_j=-i\hbar\DP{}{x^j}$$
$$\left(\frac{\partial^2\widehat f}{\partial\dot x^k\partial \dot x^\ell}\right)(0)=\left(
              \frac{\partial^2  f}{\partial x^k\partial  x^\ell}-\Gamma_{k\ell}^j\DP f{x^j}\right)(x_0). $$
In particular, for the metric tensor $T_2$ we get
$$2\,\widehat T\,f=-\hbar^2 g^{k\ell}\left(
              \frac{\partial^2 f}{\partial x^k\partial  x^\ell}-\Gamma_{k\ell}^j\DP f{x^j}\right)=-\hbar^2\Delta f,$$
where $\Delta$ is the laplacian operator associated with the metric.

For the computation of $\widehat a$ when $a$ has arbitrary order,  the formulae in \cite{Eisenhart} can be used.

\begin{obs}
The quantization is a local operation in $M$. Global conditions must be imposed once are fixed the space of admissible solutions for the wave equations.

The correspondence $\underline a\to\widehat a$ is established for the tensor as a whole and it cannot be ``factorized'' because, in the general case, is not even $\widehat{a^2}=\widehat a^2$.

Also it is  generally false that the Poisson bracket $\{\underline a,\underline b\}$ corresponds to the commutator $[\widehat a,\widehat b]$: in fact $\{\underline a,\underline b\}$ contains just first order derivatives of the coefficients of tensors $a$, $b$, while in $[\widehat a,\widehat b]$ appear, in general, derivatives whose order is the greatest order of $a$, $b$.

\end{obs}

\subsection*{Wave equations}

It is not our current topic the prolongation of operators $\widehat a$ to spaces out of $\A$.

The classical magnitudes $\underline a$ that we are considering, are functions in $TM$, which are not automatically operators on $\A$ (except when $a$ is a tensor of order 0, a function in $M$).

For each section $u\colon M\to TM$ (a vector field on $M$), the restriction $\underline a(u)$ of the function $\underline a$ to the section $u$, is transported as a function on $M$ and, as such, operates by multiplication on $\A$.

Functions $\Psi$ in $\A$, or in a prefixed space of functions or distributions, where operators $\widehat a$ and  $\underline a(u)$ coincide are the \emph{functions proper  for  the magnitude $a$ in the classical state $u$}. They are the solutions of the \emph{wave equations}
$$(\widehat a-\underline a(u))\Psi=0.$$

As an example, for a conservative mechanical system with hamiltonian $H=T+U$, the quantization of $H$ is
$$\widehat H=-\frac 12\,\hbar^2\Delta+U$$
and the wave equation is
$$\left(-\frac 12\,\hbar^2\Delta+U\right)\Psi=H(u)\Psi.$$
When $u$ is a section (a classical state of the system as a whole) of constant energy $E$, the previous equation becomes the Schrödinger equation. In particular, when $u$ is a lagrangian section, $u=\textrm{grad}\,S$, the equation $H(u)=E$ is the Hamilton-Jacobi equations for the ``action'' $S$. This equation has solutions for all the values taken for the function $H$ in the space of states. The Schrödinger equation impose on the values of the energy $E$ the condition of admitting non trivial solutions $\Psi$: it selects a subset of values admissible for the energy $H$.

\subsection*{The problem of the time evolution}
Our intutition of time is the newtonian one: time ``flows equably without relation to anything external'' (Scholium to the Definitions in the \emph{Principia}). For this reason we can accept without too much criticism the introduction of time in Quantum Mechanics rewriting the Schrödinger equation in the form $\widehat H\Phi=i\hbar\,\partial\Phi/\partial t$, with $\Phi=e^{-i\frac E\hbar t}\Psi$; the so modified ``time dependent Schrödinger equation'' admits as solutions the superposition of stationary states $e^{-i\frac {E_k}\hbar t}\Psi_k$, and the time evolution in the space of states is that of the one parametric group $\left\{e^{-i\frac t\hbar \widehat H}\right\}_{t\in\R}$.

When one takes as space of admissible ``pure'' states for the quantum-mechanic system the Hilbert space (complex and separable), Stone theorem about unitary one-parametric groups of automorphism (see \cite{Yosida}, Ch.XI, \S13) allows us to invert the direction of the transition classical $\to$ quantum: it is postulated that the evolution of the quantum system is given by a one-parametric group of unitary automorphisms; the infinitesimal generator has necessarily the form $i\times\text{(self adjoint operator)}$, and the parallelism with Classical Mechanics forces the group to be the one  generated by a constant multiple of the quantified Hamilton function (it remains the problem of find that quantified operator, a problem we take for solved). Passing the time evolution of the states to that of operators (``Heisenberg representation''), the evolution law would be $[\widehat H,\widehat a_t]=i\hbar d\widehat a_t/dt$, in analogy with the equation of the evolution for the classical magnitudes: $d\underline a/dt=\{H,\underline a\}$ (along each trajectory).

The problem is the non-existence of a function which could be called ``time'' in the formalism of the Classical Mechanics (see \cite{MecanicaMunoz}), although there exist a time of travel along each trajectory of a given mechanical system. To introduce a ``time'' function, in general it is needed the addition of a dimension to the configuration space and then to impose a constraint by a non univocally defined procedure. However, in case of conservative systems, for each lagrangian manifold of a complete integral of the Hamilton-Jacobi there is a ``time'' function that parameterizes any particular solution curve within the lagrangian manifold; except for the systems free of forces (geodesic), this time is not proportional to the  ``action'' function in the lagrangian manifold, so that front waves of the action does not move to a constant temporal rate; the ``time'' which measures the movement of the wave fronts is another one: the quotient  of action by energy. In the foundational memory of Schrödinger \cite{SchrII} the first part is dedicated to the hamiltonian analogy between Mechanics and Optics; it seems that Schrödinger is carried away by the intuition of absolute time and he use the ``t'' with two different meanings: in the formulae (3),(9), $t$ is the time that parameterizes the trajectories while in formulae (5), (6), $t$ is the quotient action/energy. With this second ``time'' the wave fronts of the action move at a constant rate.  The primitive Quantum Theory leads us to interpret the $t$ in the formulae of Quantum Mechanics in this way. On each stationary solution $\Phi=e^{-iEt/\hbar}\Psi$, the ``classical'' meaning of $t$ must be action/energy in the lagrangian manifold from which the Schrödinger equation for $\Psi$ was written.

In a subsequent paper, we will study the nature of ``time'' in Classical, Undulatory and Quantum Mechanics in conservative systems.
\bigskip


\end{document}